\documentclass[aps,prl,twocolumn,superscriptaddress,preprintnumbers,showpacs]{revtex4}
\usepackage{graphics}
\def\eq#1\en{\begin{equation} #1 \end{equation}}

\def\eqa#1\ena{\begin{eqnarray} #1 \end{eqnarray}}

\usepackage{graphicx,color}

\def\pp#1{\partial_{#1}}

\begin{document}


\title{Constraining spacetime noncommutativity with primordial nucleosynthesis}
\author{Raul Horvat}
\affiliation{Physics Division, Rudjer Bo\v skovi\' c Institute, 
Zagreb, Croatia}
\author{Josip Trampeti\'{c}}
\affiliation{Theoretical Physics Division, Rudjer Bo\v skovi\' c Institute, 
Zagreb, Croatia}

\date{\today}

\begin{abstract}
We discuss a constraint on the scale $\Lambda_{\rm NC}$ of noncommutative (NC) gauge field 
theory arising from consideration of the big bang nucleosynthesis (BBN) of light elements. 
The propagation of neutrinos in the NC background 
described by an antisymmetric tensor $\theta^{\mu\nu}$ does result in a 
tree-level vector-like coupling to photons in a generation-independent manner, raising thus a
possibility to have an appreciable contribution of 
three light right-handed (RH) fields to the energy density of the universe at nucleosynthesis
time. Considering elastic scattering processes of the RH
neutrinos off charged plasma constituents 
at a given cosmological epoch, we obtain for a conservative limit on 
an effective number of additional doublet neutrinos, $\Delta N_\nu =1$, a bound 
$\Lambda_{\rm NC}\;\stackrel{>}{\sim}$  3 TeV. With a more stringent requirement, 
$\Delta N_\nu \lesssim 0.2$, the bound is considerably improved, 
$\Lambda_{\rm NC}\;\stackrel{>}{\sim}\;\,10^3$ TeV. 
For our bounds the $\theta$-expansion of the NC action stays always meaningful, 
since the decoupling temperature of the RH species
is perseveringly much less than the inferred bound for the scale of noncommutativity. 

\end{abstract}

\maketitle


Our main motivation to discuss a constraint on the scale of noncommutative (NC) gauge field 
theory ~\cite{Madore:2000en}, $\Lambda_{\rm NC}$, arising from 
the big bang nucleosynthesis (BBN) of light elements is
physically first due to the possibility of a direct coupling of 
neutral particles to gauge bosons in the NC background 
\cite{Schupp:2002up}.
Such a background plays the role of an external field in the theory.
It will imply only a vector-like coupling between photons and neutrinos in a
U(1) gauge-invariant way, thus
endowing otherwise sterile right-handed (RH) neutrino components a new interaction.
This makes the early universe nucleosynthesis a perfect arena to set a constraint
for this interaction, giving us in turn a bound on $\Lambda_{\rm NC}$ 
if neutrinos are Dirac particles.
The coupling of matter fields to Abelian gauge bosons is a NC
analogue of the usual minimal coupling scheme.
It is possible to extend the model \cite{Schupp:2002up} to the NC electroweak
model based on the other gauge groups \cite{Calmet:2001na,Aschieri:2002mc}. 
Second reason is that the NC version of standard model (SM) \cite{Calmet:2001na} 
appears to be anomaly free \cite{Martin:2002nr}
and it does have remarkable well behaved one-loop quantum corrections \cite{Buric:2006wm},
as well as some other models do \cite{Bichl:2001cq,Latas:2007eu}. 
Third is that such models \cite{Madore:2000en}-\cite{Aschieri:2002mc}, 
up to first order in $\theta$-expansion and up to one-loop, 
do not lead to a violation of a dispersion relation 
for a quanta propagating in the vacuum 
\cite{Bichl:2001cq,Ettefaghi:2007zz}; 
they however do break Lorentz symmetry.

The NC gauge theory that we use belongs to a 
class of models that expand the NC action in $\theta$ \emph{before}
quantization~\cite{Madore:2000en}. 
These models differs fundamentally from other approaches \cite{Chaichian:2001py,Martin:1999aq,Abel:2006wj}
based on star($\star$)-products that are not $\theta$-expanded and do not use
the enveloping algebra and the Seiberg-Witten (SW) map \cite{Seiberg:1999vs}. 
Note here that the $\star$-product, a non-local bilinear expression in
the fields and their derivatives, takes the form of a series 
in antisymmetric tensor $\theta^{\mu\nu}$. The models very nicely capture 
new interactions and violations of spacetime symmetries induced by noncommutativity. 

In the model of ref. \cite{Schupp:2002up} Seiberg--Witten maps are necessary to express
the noncommutative fields $\widehat \psi$, $\widehat A_\mu$,  
(and their derivatives) that appear in the action
and transform under NC gauge transformations, in terms
of their asymptotic commutative counterparts $\psi$ and $A_\mu$.

So, we choose a perturbative approach to the NC gauge field theory in which the action 
is expanded in powers of a Poisson 
tensor $\theta^{\mu\nu}$ defined via $\star$-commutator, 
$x^\mu \star x^\nu - x^\nu \star x^\mu =ih\theta^{\mu\nu}$.
Here, the noncommutative deformation parameter $h=1/\Lambda^2_{\rm NC}$ fix 
the NC scale, while the $\theta^{\mu\nu}$ is dimensionless tensor 
with the square of matrix elements of order one.

That model is meant to provide an effective description of spacetime noncommutativity
involving the photon--neutrino contact interaction \cite{Schupp:2002up}. 
It describes the scattering of particles 
that enter from an asymptotically commutative
region into the NC interaction region.

Neutrinos do not carry an electromagnetic charge
and hence do not directly couple to photons, at least not in a commutative setting. 
In the presence of spacetime
noncommutativity, it is, however, possible to couple neutral particles to 
photons via a $\star$-commutator. The relevant NC covariant derivative 
expanded to the lowest order read 
\begin{eqnarray}
\widehat D_\mu \widehat\psi&=&\partial_\mu \widehat\psi - i\kappa e \widehat A_\mu \star \widehat\psi
+ i \kappa e \widehat\psi \star \widehat A_\mu  
\label{ncd}\\
&=&\partial_\mu \widehat \psi +  
\kappa e h\theta^{\nu\rho} \, \partial_\nu\widehat A_\mu \, \partial_\rho \widehat \psi\,, 
\nonumber
\end{eqnarray}
with the $\star$-product and 
a coupling $\kappa e$ that corresponds to a multiple (or fraction) $\kappa$ of the
electric charge $e$.  
 
To the order considered in this report, $\kappa$ can be absorbed in a rescaling of
the deformation parameter $h$. 
The $\star$-product  is associative 
but, in general, not commutative; otherwise the proposed coupling to the NC
photon field $\widehat A_\mu$ would of course be zero.

In the first equation of (\ref{ncd}), one may think of the NC
neutrino field $\widehat \psi$ as having left charge $+\kappa e$, right charge $-\kappa e$
and total charge zero. From the perspective of non-Abelian gauge theory,
one could also say that the neutrino field is charged in a noncommutative
analogue of the adjoint representation with the matrix multiplication replaced by the $\star$-product. 
For this model where only the neutrino has dual left and right charges,
$\kappa = 1$ is required by the gauge invariance of the action.

From a geometric point of view, photons
do not directly couple to the ``bare'' commutative neutrino fields,
but rather modify the NC background.

Concerning the physics to be investigated, the picture
that we have in mind is that of a spacetime that has a continuous 
`commutative' description at low energies and long distances, but a NC
structure at high energies and short distances. 
At high energies we can model spacetime using $\star$-products.
This description is expected not to be valid at low energies. 
The technical consequence is that we expand up to a certain order in $\theta$
and consider renormalization of this truncated theory up to the same order in $\theta$
\cite{Bichl:2001cq,Buric:2006wm,Latas:2007eu}.
It is obvious that in this truncated theory no UV/IR mixing effects would appear \cite{Martin:1999aq}.
This reflects very well our assumption: 
At low energies and large distances the noncommutative theory has to be modified.


The action for a neutral fermion that couples to an Abelian gauge boson 
in the NC background is:
\begin{equation}
S = \int d^4 x \left(\,\overline{\widehat\psi} \star 
i\gamma^\mu\widehat D_\mu \widehat\psi
-m \overline{\widehat\psi} \star \widehat\psi\right)\,.
\label{NCa}
\end{equation}
Here $\widehat \psi = \psi  +
e h\theta^{\nu\rho} A_\rho \pp\nu \psi$ is the NC ``chiral'' fermion field 
expanded by the ``chiral'' SW map and 
$\widehat A_\mu = A_\mu + eh\theta^{\rho\nu}A_{\nu}
\left[\partial_{\rho}A_{\mu}-\frac{1}{2}\partial_{\mu}A_{\rho}\right]$ 
is the Abelian NC gauge potential expanded by the ordinary SW map. 
Note that a ``chiral'' Seiberg-Witten map is compatible with grand unified models 
where fermion multiplets are chiral \cite{Aschieri:2002mc}.
The above action is gauge-invariant under the noncommutative U(1)-gauge transformations.

After the SW and the $\star$-product expansions we obtain 
the following electromagnetically gauge-invariant 
action at order $h$ for photons and neutrinos in terms of 
commutative fields and parameters:
\begin{eqnarray}
S &=&\int d^4 x  \bar \psi 
\Big[\frac{}{}\left(i\gamma^\mu \pp\mu  - m\right) 
\nonumber\\
&&\phantom{+ ie \theta xxxx}
-\frac{e}{2}hA_{\mu\nu}\left(
i \theta^{\mu\nu\rho}\pp\rho -\theta^{\mu\nu}m\right)\Big]\psi\;,
\label{action}
\end{eqnarray}
with ${\theta}^{\mu\nu\rho}=
{\theta}^{\mu\nu}\gamma^{\rho}+{\theta}^{\nu\rho}\gamma^{\mu}+
{\theta}^{\rho\mu}\gamma^{\nu}$.
It is interesting to note that for the massless case, Eq. (\ref{action}) 
reduces to the coupling between the stress--energy tensor of a neutrino
$T^{\mu\nu}$ and the symmetric tensor composed 
of $\theta^{\mu\nu}$ and $A_{\mu\nu}$.
This nicely illustrates our assertion that we are 
seeing the interaction of the neutrino with a modified photon--$\theta$ background.
In the above we have $A_{\mu\nu}=\partial_\mu A_\nu-\partial_\nu A_\mu$.

From (\ref{action}) we extract the following Feynman rule for massless left(right)-handed neutrinos
(the same for each generation) \cite{Schupp:2002up}: 
\begin{eqnarray}
{\Gamma}^{\mu}_{\rm L \choose \rm R}({\nu}(k){\nu}(k'){\gamma}(q))
&=&ieh\frac{1}{2}(1 \mp \gamma_5){\theta}^{\mu\nu\rho}k_{\nu}q_{\rho}\;,
\label{FR}
\end{eqnarray}
where $k = k' + q$.
Before starting computation of relevant Feynman diagrams some important comments are in order. 
When calculating S-matrix elements, one ordinarily uses the LSZ formalism, which is constructed by 
using Lorentz invariance and  implies that S-matrix elements are obtained by identifying 
the poles in momentum space of some Green functions. This happens at points satisfying 
the Lorentz invariant constraint $p^2=m^2$.
In NC spacetime Lorentz invariance is broken, which means that the poles of 
Green functions need not satisfy the constraint $p^2=m^2$; a $\theta$-dependence may 
appear and then one should be more careful when computing S-matrix elements
\footnote{However, the NC quantum correction to the photon self-energy, 
i.e. the effects of birefringence, 
as well as NC modifications of the neutrino
propagator, start to appear as late as at the 2nd order in $\theta$-expansion
\cite{Bichl:2001cq,Abel:2006wj,Ettefaghi:2007zz}. 
So, to first order in $\theta$-expansion, the one-loop quantum correction 
shows that the photon and neutrino propagators still have poles of the ordinary type.}.
Since we have to compute only S-matrix elements (amplitudes) 
at tree level and only up to first order in $\theta$,
the usual way of deriving matrix elements remains valid.

Signatures of noncommutativity and/or the bounds on the NC scale come from
neutrino astrophysics
\cite{Schupp:2002up,Minkowski:2003jg,Haghighat:2009pv}, high energy particle
physics \cite{Melic:2005hb,Buric:2007qx,Alboteanu:2006hh},
as well as from other low energy nonaccelerator experiments \cite{Carroll:2001ws}.
The bound from $Z \rightarrow \gamma\gamma$ decay on $\Lambda_{\rm NC}$,
of order a few TeV \cite{Buric:2007qx}, is the most robust due to the finite one-loop
quantum corrections of the gauge sector of the nmNCSM \cite{Buric:2006wm}.
In the following we are going to place, for the first time, a bound on
the spacetime noncommutativity from cosmology, 
by making use of the BBN restriction on the number 
of neutrino species.

For several past decades, the BBN has established itself as one of the most
powerful available probes of physics beyond the SM, giving
many interesting constraints on particle properties (an extensive summary is
available, for instance, in \cite{Sarkar}). The BBN had begun to play a
central role in constraining particle properties since the seminal paper of
Steigman, Schramm and Gunn \cite{Steigman:1977kc}, in which the 
observation-based determination of the primordial abundance of $^{4}$He
was used for the first time to constrain the number of light
neutrino species. Later, with the inclusion of other light element abundances
(D, $^{3}$He and $^{7}$Li) and their successful agreement with the
theoretically predicted abundances, many aspects of physics beyond SM can
have been further constrained \cite{Malaney:1993ah}. 

One uses to parametrize the energy density of new relativistic particles 
at the time of BBN in
terms of the effective additional number of neutrino species, $\Delta
N_{\nu}$, whose determination requires both a lower limit to
the barion-to-photon ratio ($\eta \equiv n_b/n_{\gamma }$) and an upper bound 
for the primordial mass fraction of $^{4}$He, $Y_p $ \cite{Olive:1980bu}. By 
increasing $\Delta
N_{\nu}$ increases the Hubble expansion rate, causing higher freeze-out
temperature for the weak interaction reactions 
and therefore larger neutron fraction, resulting eventually in a larger value
of $Y_p $. With the WMAP value 
for $\eta $ \cite{22}, $Y_p $ was predicted to increase \cite{Cyburt:2004cq}, having a 
tendency  to
loosen the tight bounds on $\Delta N_{\nu}$ that existed before. With a 
recent
reanalysis \cite{Cyburt:2004yc} of the observed primordial $^{4}$He abundance, and its
uncertainties, even the higher mean value for $Y_p $ was predicted (0.2495).
Consequently, the limits to $\Delta N_{\nu}$ get relaxed. Therefore, in our
analysis $\Delta N_{\nu} = 1$ will be employed as a central value 
to constrain nonstandard
physics, a marginally conservative limit for the assessment \cite{Cyburt:2004yc}.      

The energy density of 3 light RH neutrinos during the BBN is equivalent to the
effective number $\Delta N_{\nu}$ of additional doublet neutrinos
\begin{equation}
\Delta N_{\nu} = \sum_{i = 1}^{3} \frac{g_{i}}{2}
\left (\frac{T_{\nu_{R}}}{T_{\nu_{L}}} \right )^4 \;,
\label{7}
\end{equation} 
where the number of spin degrees of freedom $g_{i} = 2$ and $T_{\nu_{L}}$
is the temperature of the SM neutrinos, being the same as
that of photons down to $T \sim 1$ MeV. Hence we have
\begin{equation}
3 \left (\frac{T_{\nu_{R}}}{T_{\nu_{L}}} \right )^4 \lesssim \Delta N_{\nu,
max} \;.
\label{8}
\end{equation} 

How the temperature of $\nu_{R}$s,
decoupled at $T_{dec}$, relates to the temperature of 
still interacting $\nu_{L}$s below
$T_{dec}$, stems easily from the fact that the entropy in the decoupled species
and the entropy in the still interacting ones are separately conserved. The
ratio of the temperatures is a function of $T_{dec}$, and is given by
\cite{Srednicki:1988ce} \footnote{Note that while $g_{*\nu_R }$ does not change following
decoupling, which is certainly the case for light neutrinos in our
consideration, the ratio (\ref{9}) acquires a more familiar form
$\frac{T_{\nu_{R}}}{T_{\nu_{L}}} = \left [
\frac{g_{*S}(T_{\nu_L})}{g_{*S}(T_{dec})} \right ]^{1/3}$ .}
\begin{equation}
\frac{T_{\nu_{R}}}{T_{\nu_{L}}} = \left [\frac{g_{*\nu_R
}(T_{dec})}{g_{*\nu_R }(T_{\nu_L})}
\frac{g_{*S}(T_{\nu_L})}{g_{*S}(T_{dec})} \right ]^{1/3} \;,
\label{9}
\end{equation}
where $g_{*\nu_R }$ and $g_{*S}$ are the degrees of freedom specifying the
entropy of the decoupled and of the interacting species, respectively \cite{Kolb,Sarkar}. 
Now, combining (\ref{8}) with (\ref{9}) and noting that at the time of BBN
$g_{*S}(\sim MeV) = 10.75$, one arrives at
\begin{equation}
g_{*S}(T_{dec})\; \stackrel{>}{\sim} \; \frac{24.5}{(\Delta N_{\nu,max})^{3/4}} \;.
\label{10}
\end{equation}
With the conservative bound $\Delta N_{\nu,max} =1$, (\ref{10}) implies
$g_{*S}(T_{dec}) > 24.5$ which in turn enforces $T_{dec} > T_C $, where $T_C
$ is the critical temperature for the deconfinement restoration phase
transition, $T_C \sim 200$ MeV.

Using the nonstandard $\bar\nu \nu \gamma $ vertex as given by (\ref{FR}), the cross
section for scattering of the RH neutrino off a charged fermion (with a
charge e) in the early universe is calculated to be  
\begin{equation}
\sigma_{scatt} \simeq 36 \; \alpha^2 \frac{E^2}{\Lambda_{\rm NC}^4} \;,
\label{11}
\end{equation}
where $\alpha $ is the fine-structure constant. A remark is in order
regarding arrival to (\ref{11}). The forward scattering 
gives rise to a logarithmic singularity, ln$(q_{max}/q_{min})$, which we
regularize by using a Debye mass, $q_{min} = m_{el} \simeq eT/\sqrt{3}\,.$
Also, the mass of the charged fermion is neglected in (\ref{11}).

The RH neutrino is commonly considered to decouple at the temperature 
$T_{dec}$ when the condition
\begin{equation}
\Gamma_{scatt} (T_{dec}) \simeq H(T_{dec}) \;,
\label{12}
\end{equation}
is satisfied.  
Here $\Gamma_{scatt} = \; < n_{scatt}\;\sigma_{scatt}\;v >$ is the 
thermally-averaged scattering rate, and  
expansion rate of the universe in the
radiation-dominated epoch is given by
\begin{equation}
H \simeq 1.66 \; g_{*}^{1/2} \frac{T^2 }{M_{Pl}} \;,
\label{13}
\end{equation}
where $T$ denotes a common temperature for the interacting species.
Here $g_{*}$ counts degrees of freedom specifying the energy density and
coincides with $g_{*S}$ always when particle species have a common
temperature. 

Using $n_{scatt} \simeq 0.18 \; T^3$, $E \;\simeq \; 9\;T$ and $\sigma_{scatt}$ from (\ref{11}), 
we obtain the following decoupling temperature
\begin{equation}
T_{dec} \simeq 0.5 \; \alpha^{-2/3} M_{Pl}^{-1/3} \Lambda_{\rm NC}^{4/3} \;.
\label{14}
\end{equation}

Now, by imposing a constraint $T_{dec} > T_{C} \sim 200$ MeV, and counting
all charged fermions present in the cosmological epoch between $T_{C}$ and 
a neighboring  quark mass above $T_C$, we find $\Lambda_{\rm NC} \;\stackrel{>}{\sim}\; 3 $ TeV.
With the inclusion of electrons, muons and $s$ quarks, this is
our result based on the conservative bound $\Delta N_{\nu,max} =1$.

A new analysis of the $^{4}$He abundance \cite{Cyburt:2004yc}, indicating a shift
in the mean value of $Y_{p}$, quoted also a relaxed limits to $\Delta
N_{\nu}$ compared to those that had existed before (for a review see again
\cite{Sarkar}). This also diminishes a potential for BBN to constrain nonstandard
physics, reducing also our bound on $\Lambda_{\rm NC}$. One can hope that
constant effort in improving the precision of the observation of primordial
abundances of light elements will result in tighten limits to $\Delta
N_{\nu}$ once again. This would also have important consequences for our 
bound too. For instance, $\Delta N_{\nu} \lesssim \; 0.2$ would enforce $T_{dec}$
close to a critical temperature of the electroweak phase transition,
$T_{dec} \lesssim \; 300 $ GeV, implying in turn a significantly better limit,
$\Lambda_{\rm NC} \;\stackrel{>}{\sim} \; 10^3 $ TeV. This limit purports
inclusion of all the charged leptons and quarks (as seen from the Table 1. in \cite{Sarkar}).

Finally, a remark on the consistency of our approach is in order. As seen
from (\ref{FR}), the perturbative expansion in terms of the scale of
noncommutaivity $\Lambda_{\rm NC}$ retains its meaningful character only if
$E^2 /\Lambda_{\rm NC}^2 \lesssim \; 1$, where $E$ is the characteristic energy for a
process. For a cosmological setting, as given above, this translates to
$T_{dec}^2 /\Lambda_{\rm NC}^2 \lesssim \; 1$. Our bounds always fulfill this
constraint.  

In conclusion, we have demonstrated how the BBN bounds on the scale of spacetime
noncommutativity appear among the best ones even for large errors inherent
in observation of primordial abundances of light elements, which sensibly 
diminishes the potential of the method. We have also shown that  
with improving the precision of 
observation, the BBN bounds could easily  become the strongest restriction on 
the scale of noncommutativity.

\section*{Acknowledgement}
The work of R.H. and J.T. are supported by the  Croatian  Ministry of Science, Education and Sports 
under Contract Nos. 0098-0982930-2872 and 0098-0982930-2900, respectively. 
The work of J.~T. is in part supported by the EU (HEPTOOLS) project under contract MRTN-CT-2006-035505.

\end{document}